\documentclass[epj]{svjour}
\usepackage{graphics}
\newcommand{\gS}[1]{#1\!\!\!\!\!\not~}

\newcommand{\pslash}{\gS{p}}

\begin{document}

\title{Aspects of quark mass generation on a torus}

\author{C. S. Fischer\inst{1} \and M. R. Pennington\inst{2}
\thanks{Talk given by CF at QNP06, Madrid, June 2006.}}                    
\institute{Institut f\"ur Kernphysik, Technical University of Darmstadt, 
Schlossgartenstra{\ss}e 9, 64289 Darmstadt, Germany \and
IPPP, Durham University, Durham DH1 3LE, U.K.}

\date{Received: date / Revised version: date}

\abstract{In this talk we report on recent results for the quark 
propagator on a compact manifold. The corresponding Dyson-Schwinger 
equations on a torus are solved on volumes similar to the ones 
used in lattice calculations. The quark-gluon interaction is fixed 
such that the lattice results are reproduced. We discuss both the
effects in the infinite volume/continuum limit as well as effects 
when the volume is small.
}

\maketitle

\section{Introduction}
\label{intro}

In the low energy sector of QCD dynamical chiral symmetry breaking 
is the dominant nonperturbative effect with respect to hadron 
phenomenology. Thus any framework attempting to explain the spectra
and decay properties of light hadrons has to satisfy all constraints 
related to the correct pattern of chiral symmetry breaking in QCD. 
In this respect lattice simulations have to deal with two types of
problems: on the one hand, commutation relations bet\-ween $\gamma$-matrices 
are affected by the finite lattice spacing. On the other hand, there
is the finite volume. Since continuous symmetries cannot be 
spontaneously broken at a finite volume $V$, chiral symmetry is restored 
in the limit of zero current quark mass, $m \rightarrow 0$ (see {\it e.g.}
\cite{Leutwyler:1992yt}). Thus one first has to perform both the 
continuum and infinite volume limit before one can investigate the 
chiral limit. In turn, studies with physical up- and down-quark masses 
necessarily require large volumes to avoid chiral restoration effects.

It is certainly desirable to study volume effects in chiral symmetry 
breaking in different frameworks. Chiral perturbation theory has turned 
out to be a reliable tool for both volume and chiral extrapolations.
On the other hand, chiral perturbation theory has nothing to say about 
volume effects in the underlying quark and gluon substructure. For this 
the Green's function approach employing Dyson-Schwinger equations (DSEs) 
\cite{Alkofer:2000wg,Fischer:2006ub} provides a suitable alternative. 
In this talk we present results for the quark propagator in Landau gauge 
QCD. Based on ideas developed in \cite{Bhagwat:2003vw,Fischer:2005nf}, 
we use lattice results to fix the quark-gluon interaction at quark masses 
accessible on the lattice. To this end we solve the DSEs on
tori with similar volumes. We then change current quark masses and volumes 
to explore the corresponding effects on the quark propagator.

\section{Quark-gluon interaction}
\label{sec:1}

\begin{figure}[t]
\resizebox{0.4\textwidth}{!}{\includegraphics{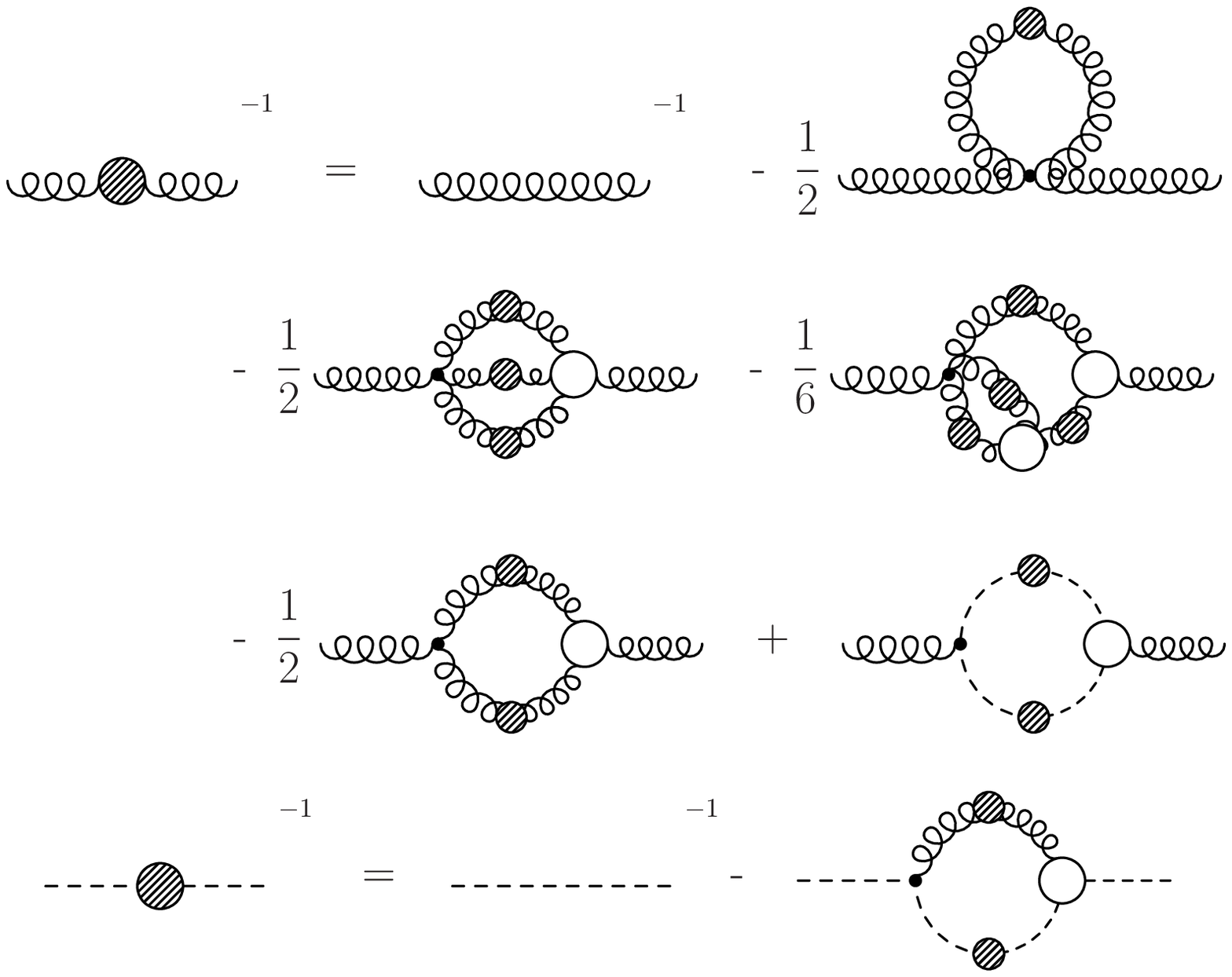}}
\resizebox{0.38\textwidth}{!}{\includegraphics{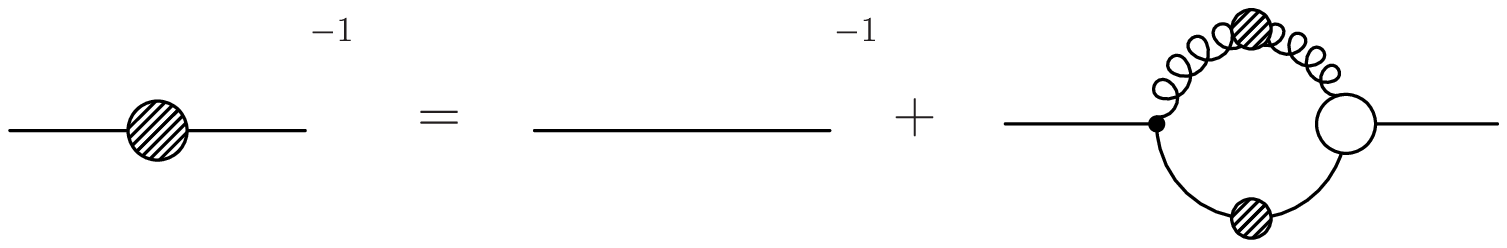}}
  \caption{Dyson-Schwinger equations for the gluon (curly), 
         ghost (dashed) and quark (straight lines)
           propagators. Filled circles denote dressed propagators 
	   and empty circles denote dressed vertex functions.}
	   \label{fig:ghostglue}
\end{figure}

The Dyson-Schwinger equations (DSEs) for the ghost, gluon and quark propagators
$D_G, D_{\mu,\nu},S(p)$ 
\begin{eqnarray}
D_{G}(p)       &=& - \frac {G(p^2)}{p^2} \;,                     \label{Gh-prop}\\
D_{\mu \nu}(p) &=& \left(\delta_{\mu \nu} - \frac{p_\mu p_\nu}{p^2} \right) 
                   \frac{Z(p^2)}{p^2} \;,                        \label{Gl-prop}\\
S(p)           &=& \frac{Z_f(p^2)}{i \pslash + M(p^2)} \;,       \label{Qu-prop} 
\end{eqnarray}
in quenched approximation are given diagrammatically in Fig.~\ref{fig:ghostglue}. 
This system of equations is closed but for the dressed vertices, which have to be 
approximated by suitable ans\"atze. The truncation scheme for the Yang-Mills sector 
is discussed {\it e.g.} in \cite{Fischer:2006ub,Fischer:2002hn}. The corresponding 
numerical solutions for the ghost and gluon propagator can be represented accurately by
\begin{eqnarray}
R(p^2) &=& \frac{c \,(p^2/\Lambda^2_{YM})^{\kappa}+d \,(p^2/\Lambda^2_{YM})^{2\kappa}}
{1+ c \,(p^2/\Lambda^2_{YM})^{\kappa}+d \,(p^2/\Lambda^2_{YM})^{2\kappa}} \,, \nonumber\\
Z(p^2) &=& \left( \frac{\alpha(p^2)}{\alpha(\mu^2)} \right)^{1+2\delta} R^2(p^2) \,, \label{glue}\\ 
G(p^2) &=& \left( \frac{\alpha(p^2)}{\alpha(\mu^2)} \right)^{-\delta} R^{-1}(p^2)  \,,
\label{ghost}
\end{eqnarray}
with the scale $\Lambda_{YM}=0.658 \,\mbox{GeV}$, the coupling $\alpha(\mu^2)=0.97$ and 
the parameters $c=1.269$ and $d=2.105$ in the auxiliary function $R(p^2)$. The quenched 
anomalous dimension of the ghost is given by $\delta=-9/44$ and related via $\gamma=-1-2\delta$
to the one of the gluon. The infrared exponent $\kappa$ is determined in an analytical
infrared analysis \cite{Lerche:2002ep}: $\kappa = (93-\sqrt{1201})/98 \approx 0.595$. 
The running coupling $\alpha(p^2)$, defined via the nonperturbative ghost-gluon vertex
$\alpha(p^2) = \alpha(\mu^2)\, G^2(p^2)\, Z(p^2)$, can be represented by
\begin{eqnarray}
\alpha(p^2) &=& \frac{1}{1+p^2/\Lambda^2_{YM}} 
\bigg[\alpha(0) + p^2/\Lambda^2_{YM} \times \nonumber\\
&& \hspace*{-2mm}\left.\frac{4 \pi}{\beta_0} \left(\frac{1}{\ln(p^2/\Lambda^2_{YM})}
- \frac{1}{p^2/\Lambda_{YM}^2 -1}\right) \right]\,.  
\end{eqnarray}
where $\alpha(0) \approx 8.915/N_c$ is also known analytically \cite{Lerche:2002ep}.

Apart from the gluon propagator the other missing piece in the quark-DSE is the quark-gluon
vertex. We use an ansatz of the form \cite{Fischer:2005nf}
\begin{equation}
\Gamma_\nu(k,\mu^2) \;=\; \gamma_\nu\, \Gamma_{1}(k^2)\, \Gamma_{2}(k^2,\mu^2)\, \Gamma_{3}(k^2,\mu^2) \label{v1}
\end{equation}
with the components
\begin{eqnarray}
\Gamma_{1}(k^2) &=&  \frac{\pi \gamma_m}{\ln(k^2/\Lambda_{QCD}^2 +\tau)}\,, \label{v2}\\[3mm]
\Gamma_{2}(k^2,\mu^2) &=& G(k^2,\mu^2)\ G(\zeta^2,\mu^2)\ \widetilde{Z}_3(\mu^2) \nonumber\\[-2.mm]
&& \\[-2.mm]
&&\hspace{10mm} \times \ h \ [\ln(k^2/\Lambda_{g}^2 +\tau)]^{1+\delta} \label{v3}\nonumber\\[3mm]
\Gamma_{3}(k^2,\mu^2) &=& Z_2(\mu^2)\; \frac{a(M)+k^2/\Lambda_{QCD}^2}{1+k^2/\Lambda_{QCD}^2}\,, \label{v4}
\end{eqnarray}
where $\gamma_m=12/33$ is the quenched anomalous dimension of the quark and 
$\tau = e-1$ acts as a convenient infrared cutoff for the logarithms. The 
quark mass dependence of the vertex is parametrised by
\begin{equation}
a(M)\; =\; \frac{a_1}{1 + a_2 M(\zeta^2)/\Lambda_{QCD} + a_3 M^2(\zeta^2)/\Lambda_{QCD}^2},
\label{am}
\end{equation}
where $M(\zeta^2)$ is determined during the iteration process at $\zeta=2.9$ GeV.

In \cite{Fischer:2005nf} the parameters have been fitted such that lattice results for the
quark propagator \cite{Bowman:2002bm,Zhang:2004gv} are reproduced on similar compact manifolds.
To this end, solutions for the ghost and gluon propagators on similar manifolds have been used
\cite{Fischer:2005ui}. These solutions have been improved in \cite{Fischer} with the 
effect that the correct infinite volume limits can be seen on tori with volumes larger than 
$V \approx (10 {\rm fm})^4$. Here we use the improved solutions to update the results for the 
quark propagator reported in \cite{Fischer:2005nf}. The resulting modified values for the parameters
(staggered quarks \cite{Bowman:2002bm} only) are given in Fig.~\ref{tab}. The numerical results,
shown below, are discussed in the next section.
 
\begin{table}
\begin{center}
\begin{tabular}{|c||c|c|c|c|c|c|}
\hline    &  h    &  $\Lambda_g$  & $\Lambda_{QCD}$ &  $a_1$  &  $a_2$ & $a_3$ \rule[-1mm]{0mm}{5mm}\\
          &       &     (GeV)     &     (GeV)       &         &        &        \\ \hline\hline
staggered &  1.48 &      1.50     &      0.35       &   18.39 &  5.25  & -0.21 \rule[-3mm]{0mm}{7mm} \\\hline
\end{tabular}
\caption{Parameters used in the vertex model, Eqs.~(\ref{v1}-\ref{v4}).\label{tab}}
\end{center}
\end{table}

\begin{figure}[b]
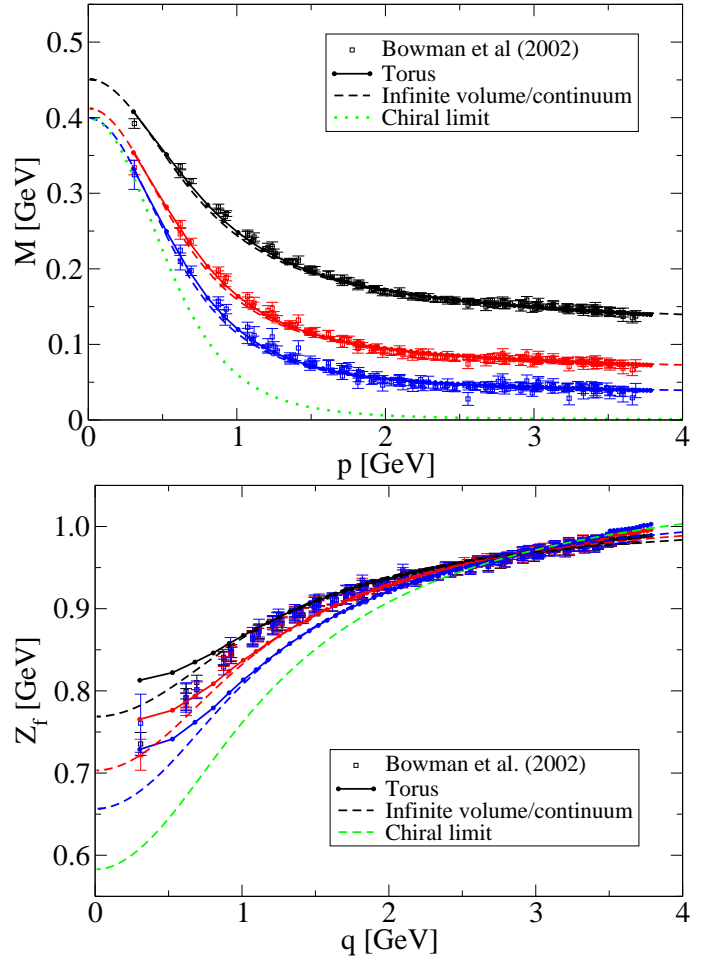

\resizebox{0.5\textwidth}{!}{\includegraphics{tor_latt.eps}}
\resizebox{0.5\textwidth}{!}{\includegraphics{tor_latt_Z.eps}}
  \caption{The quark mass function $M(p^2)$ and the wave function $Z_f(p^2)$ for renormalised current 
  quark masses of $m(2.9 \,{\rm GeV})=44, \, 80,\, 151 \,{\rm MeV}$. Compared are lattice results of 
  \cite{Bowman:2002bm} with DSE-solutions on a torus (straight lines with filled circles) and in the 
  infinite volume/continuum limit (dashed lines).}
	   \label{fig:quark}
\end{figure}

\section{Numerical results}

Our results for the quark mass function $M(p^2)$ and the wave function $Z_f(p^2)$ for renormalised 
current quark masses of $m(2.9 \,{\rm GeV})=44, 80, 151 \,{\rm MeV}$ are displayed in Fig.~\ref{fig:quark}. 
The data for the quark mass function from the staggered lattice action can be very well reproduced
with the simple vertex ansatz specified in the last section. The results for the wave functions agree
less with each other: the lattice data for the three different masses are smaller spread than the 
DSE-results. This may be an indication for the importance of tensor structures in the quark-gluon 
vertex different from the simple $\gamma_\mu$-term used in Eqs.~(\ref{v1}-\ref{v4}). In 
Fig.~\ref{fig:quark} we also show results in the infinite volume/continuum limit. For the quark mass
function these are very close to the results on the compact manifold, indicating that the used
lattice volume, $V = (2.04 \,\,{\rm fm})^4$ is indeed large enough s.t. the correct pattern of chiral symmetry
breaking can be observed for the quark masses shown. Small effects are only seen in the renormalisation
point dependent and therefore unphysical quark wave function $Z_f(p^2)$. This result agrees with
conclusions from volume studies on the lattice \cite{Zhang}.\footnote{The larger volume effects found
in \cite{Fischer:2005nf} also for the mass function are due to an overestimation of the 
effects in the Yang-Mills sector. An improved analysis of volume effects in the ghost and gluon
propagators can be found in \cite{Fischer}.}

The resulting quark mass and wave functions in the infinite volume, continuum and chiral limit are also shown in 
Fig.~\ref{fig:quark}. The corresponding quark mass at zero momentum is $M(0)=400 \,{\rm MeV}$. In the 
complex momentum plane the propagator has a leading analytic structure given by a pair of 
conjugate poles at $m = 416(20) \pm i \, 304(20)$ MeV. The corresponding spectral function violates 
positivity and is therefore characteristical for a confined quark. The chiral condensate  
is $|\langle \bar{q}q\rangle|_{\overline{MS}}^{2 GeV} = (235 \pm 5 \,{\rm MeV})^3$,
using the quenched scale $\Lambda_{\overline{MS}}=225(21) \mbox{MeV}$ in the 
$\overline{MS}$ scheme.
The pion decay constant, as determined from solutions of the pion Bethe-Salpeter equation, is 
$f_\pi = 72 \,{\rm MeV}$. Both values are considerably smaller than the experimental result. Taken 
at face value this means that the quenched theory underestimates the amount of chiral symmetry breaking 
by about 10-20 percent. A similar conclusion has been drawn in \cite{Bhagwat:2003vw}.

\begin{figure}[t]
\resizebox{0.4\textwidth}{!}{\includegraphics{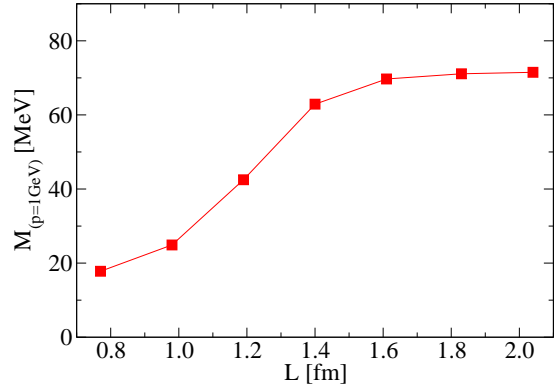}}
  \caption{The quark mass function for an up/down quark at a given momentum $p=1\mbox{GeV}$ 
plotted as a function of the box length $L=V^{1/4}$.}
	   \label{fig:M_L}
\end{figure}
Finally we study the behaviour of the quark mass function at small volumes. In Fig.~\ref{fig:M_L}
we plot $M(p=1 \,{\rm GeV})$ as a function of the box length $L$. The corresponding current 
quark mass is of the order of a typical up/down-quark mass, $M(p= 2.9 \,\mbox{GeV})=10 \,\mbox{MeV}$.
We clearly see that the quark mass function grows rapidly in the range $1.0 \, <\, L \, <\,  1.6$~fm 
signalling the onset of dynamical chiral symmetry breaking. Above $L\,=\, 1.6$~fm, a plateau is reached.
This picture does not change when we extract the mass function $M(p^2)$ at smaller momenta $p^2$ or 
when we employ even smaller quark masses. The value is also not dependent on the type of quark-gluon
interaction used. The result indicates that a safe volume to reproduce the correct pattern of chiral 
symmetry breaking on a compact manifold is at least 
\begin{equation}
L_{\chi SB}\; \simeq\; 1.6 \,\,\mbox{fm}.
\end{equation}
In particular this gives a (surprisingly small) miminal box length below which chiral perturbation 
theory cannot safely be applied. 
 
\subsection*{Acknowledgements}

We thank the organisers of QNP06 for all their efforts. CF has been supported by the 
Deutsche Forschungsgemeinschaft (DFG) under contract Fi 970/7-1.

\end{document}